\documentclass[conference]{IEEEtran}
\usepackage{cite}
\ifCLASSINFOpdf
   \usepackage[pdftex]{graphicx}
   \graphicspath{{images/}{images/}}
\else
\fi
\usepackage{amsmath}
\begin{document}
\title{Multimodal Fusion Based Attentive Networks for Sequential Music Recommendation}

% \and
% \and

\author{\IEEEauthorblockN{Kunal Vaswani}
\IEEEauthorblockA{IIIT Hyderabad\\
Hyderabad, India\\
kunal.vaswani@students.iiit.ac.in}
\and
\IEEEauthorblockN{Yudhik Agrawal}
\IEEEauthorblockA{IIIT Hyderabad\\
Hyderabad, India\\
yudhik.agrawal@research.iiit.ac.in}
\and
\IEEEauthorblockN{Vinoo Alluri}
\IEEEauthorblockA{IIIT Hyderabad\\
Hyderabad, India\\
vinoo.alluri@iiit.ac.in}}


%Telephone: (800) 555--1212, Fax: (888) 555--1212}


\maketitle

\begin{abstract}
Music has the power to evoke intense emotional experiences and regulate the mood of an individual. With the advent of online streaming services, research in music recommendation services has seen tremendous progress. Modern methods leveraging the listening histories of users for session-based song recommendations have overlooked the significance of features extracted from lyrics and acoustic content. We address the task of song prediction through multiple modalities, including tags, lyrics, and acoustic content. In this paper, we propose a novel deep learning approach by refining Attentive Neural Networks using representations derived via a Transformer model for lyrics and Variational Autoencoder for acoustic features. Our model achieves significant improvement in performance over existing state-of-the-art models using lyrical and acoustic features alone. Furthermore, we conduct a study to investigate the impact of users' psychological health on our model's performance.

\end{abstract}

\begin{IEEEkeywords}
Music Recommendations, Multimodal, Attention Networks, User Evaluation
\end{IEEEkeywords}

% \fi
\IEEEpeerreviewmaketitle

\section{Introduction}
Music subscription platforms have seen a massive increase in subscriptions amid the Covid outbreak. With the rise of isolation, music has given societies a means to cope with adversity. One of the primary areas these platforms have targeted to enhance their user engagement is Big Data. Big Music Data includes massive song libraries and the temporal engagement obtained from users' listening patterns. Researchers in music recommendations have refined their algorithms to create a more compelling user experience thanks to the enormous amount of data produced by music streaming services. Recommendation systems (RS) play a critical role in improving the user experience and increasing user growth in these platforms. Collaborative Filtering (CF) and Content-Based (CB) approaches have gained prominence in music RS. CF techniques utilized in the past rely on combining users' preferences by using user-item interactions \cite{ringo} but suffer from cold-start problems like working with new inputs. On the other hand, CB methods have gained popularity in recent years since they use track-based content for suggestions and cope with cold-start problems. Music possesses representation in predominantly three modalities, including acoustic content, user-defined tags comprising song/track descriptors (ex: artist, genre, mood, instruments), and lyrics. Music RS has been developed in the field of Music Information Retrieval (MIR) research, typically utilizing information from either an individual modality (ex: only tags) \cite{sachdevaetal} or at most two modalities (ex: acoustic features + text) \cite{oramasetal}. However, scarce studies have been done in creating a multimodal system that capitalizes on joint information across modalities.

The majority of Music RS have relied on acoustic features for their design \cite{acoustics1,acoustics2}. Acoustic features also play a vital role in delivering enriched musical representations as shown in studies like emotion recognition \cite{Rocha2014}. Furthermore, users' preferred musical genres' acoustic characteristics are known to influence the songs they download in their non-preferred musical forms \cite{baroneetal}. Emotions and user traits are essential factors in musical tastes and hence necessary for music recommendations \cite{peia,andersonetal,ferwerdaetal}. We provide a novel method for incorporating acoustic features into downstream tasks by generating latent representations with a variational autoencoder. Recent MIR studies based on lyrics in extracting emotions \cite{lyrics2emotions} have revealed the depth of knowledge embedded in lyrics. Advanced neural techniques like transformers that do not limit themselves to extracting shorter patterns in lyrical data are the critical reason for progress in these approaches. These have been significantly under-explored in music RS, we approach this by embedding information from lyrics using sentence transformers. In addition to lyrics and acoustic features, another modality represented by user-defined tags has provided a means to consolidate user interests in Music RS \cite{sachdevaetal}. Surana et al. \cite{surana2020tag2risk,suranaetal} demonstrate how user-specific states, represented by psychological well-being, modulate musical choices characterized by tags and acoustic features on online streaming platforms. However, as mentioned before, there is a dearth of studies combining these modalities to model user behavior to create highly personalized Music RS.

A key step in the design of a Music RS is to dynamically predict the user's current song preferences based on previous listening history. Aside from the multimodal characteristics that can be used to make a Music RS, integrating users' temporal engagement is a challenging task in itself. Deep learning algorithms have become influential in predicting songs based on a user's listening history. Although other architectures, such as CNNs, have been shown to do well in deep learning tasks, we prefer to use Attentive Networks with Recurrent Neural Units because of their superior ability to collect sequential information. With the increase in modalities for our task, we chose the path of multimodal fusion \cite{fusion1, fusion2}, where individual models first focus on extracting features from individual modalities to be subsequently combined via a deep learning architecture to predict song preferences. To demonstrate the effect of individual states on the performance of our model, for the first time, we perform a case study by evaluating our developed model on users categorized to be at risk for depression. This work thereby also highlights the importance of incorporating individual differences in future Music RS. We evaluate the effectiveness of our approach by using a large dataset from Last.fm\footnote{https://www.last.fm/}, which included 413k unique songs from 541 users, and compare it with state-of-the-art models. 

\section{Related Work}
Collaborative recommendation techniques rely on users' ratings of various tracks in the system. The cold start problem is one of the primary challenges that early collaborative systems \cite{Smith} in music recommendations encounter. When a new track is uploaded to the system, the algorithm finds it challenging to propose it to users as there are fewer interactions between users and the track. The problem becomes incredibly difficult with big platforms like Spotify, with over 70 million tracks\footnote{https://newsroom.spotify.com/company-info/} and users interacting with only a tiny percentage of them. Owing to the cold-start problems faced by collaborative filtering, content-based approaches have gained popularity in music RS. These approaches use the data from music content to produce latent item vectors, which are then used in collaborative filtering methods or to create hybrid models. For example, convolutional neural networks were used by van den Oord et al. \cite{vandenoordetal} to predict latent factors from audio signals and compared with traditional bag-of-words models. CNNs in their experiments show superior performance on a music recommendation task, thus demonstrating the advantage of the latest deep learning techniques over traditional approaches. Currently, one commonly used approach to get acoustic features is to use Spotify's API\footnote{https://developer.spotify.com/}. 
%and is commonly used in many studies. 
For example, Zangerle et al. \cite{acoustics1} integrate Spotify audio features in their model to describe the musical preferences of users. Audio features have also been used in conjunction with other modalities; Oramas et al. \cite{oramasetal} use CNNs to create track embeddings from audio signals and fuse them with artist embeddings created from artist biographies. They show performance improvement compared to using only artist embeddings or track embeddings, demonstrating the importance of fusing modalities.

The fast-growing advancements in Natural Language Processing techniques to create representations of lyrics and other text data (ex: user reviews, artist biographies) are gaining importance. For example, Lin et al. \cite{linetal} use textual embeddings in their architecture, but their approach is limited to traditional paragraph vectors \cite{doc2vec}. Similarly, Gossi and Gunes \cite{Gossi2016} use early TF-IDF methods to model lyrical data. Vystrčilová and Peška \cite{lyricsoraudio} provide a comparative study on the usage of various NLP techniques to capture lyrical embeddings. Though Vystrčilová and Peška use some of the latest techniques for extracting lyrical features, they do not use deep learning architectures for song predictions. Nonetheless, advanced NLP techniques like transformer models \cite{bert,vaswanietal} that can catch long-term dependencies in text, make them better suited to handling lyrical data, and remain unexplored in deep learning-based music RS. 

We tend to listen to music in a certain order interspersed with periods of inactivity. The listening periods of users surrounded by their periods of inactivity are called sessions. The user's states and traits influence these patterns. For example, individuals scoring high on trait neuroticism, also characterized by their high psychologically distressed states, have been found to engage in repetitive listening patterns across sessions \cite{suranaetal}. Hence, the temporal evolution of music consumption captures relevant user-specific preferences. Sequence-aware RS is an apt choice for this since it incorporates temporal dependencies and benefits from session-based patterns. Session-based approaches have been used to model sequential data, which work by feeding user clicks into RNNs, generating predictions based on the users' previous clicks  \cite{gru4rec}. Furthermore, it is recognized that Attentive Neural Nets enhance efficiency over these by paying attention to the user's session history; Attention is applied to the RNN outputs produced by sessions. These were observed in the experiments conducted by Sachdeva et al. \cite{sachdevaetal}, and Lin et al. \cite{linetal}. Sachdeva et al. focus only on tag modalities and use one-hot encoded inputs for the same. They generate tag representations all while performing song predictions and train their architecture end-to-end. This poses difficulties when working with Big Data and limits the scalability of the approach. Experiments by Lin et al. provide a direction to resolve these issues using latent representations from graphic, textual, and visual data. However, they do not use more relevant data for extracting song features such as acoustic content. Also, we demonstrate how advanced techniques can be used to generate representations for the modalities they utilize, such as lyrics.

Based on the existing research mentioned above, we focus on utilizing various song representations using a multimodal fusion approach. First, we leverage sessions instead of using users' entire listening history to generate sequential representations. Further, we take advantage of individual modalities, such as acoustic features and undervalued lyrics, to show their impact and conduct experiments on song predictions by combining them with sequential representations. Eventually, we fuse information coming from all the modalities to obtain a multimodal architecture. We also present a comparative study in which we demonstrate the value of information from each model and whether any complementary information exists that could benefit the fusion of these models. Finally, since user states modulate musical choices, we further evaluate our model on two groups of individuals: the ones at risk (At-Risk) for depression and those not at risk (No-Risk). Owing to the aforementioned repetitive listening patterns exhibited by at-risk individuals, we predict that our model would demonstrate higher prediction accuracy for them when compared to the No-Risk group.

%-----------------------------------------------------------------
%-----------------------------------------------------------------
\section{Method}
\label{section:method}

\subsection{Dataset}
Data was obtained from a previous study \cite{surana2020tag2risk} comprising listening histories of 541 Last.fm users (82 females, mean age = 25.4 years, std = 7.3 years). Most of them belonged to the United States and the United Kingdom accounting for about 30\% and 10\% of the participants respectively. Every other country contributed to less than 5\% of the total participants. The participants' listening histories were extracted for a duration of 6 months extracted around the time they took part in the survey which comprised collecting their well-being scores. The respective users' well-being was assessed using a standard diagnostic questionnaire (Kessler's Psychological Distress Score or K-10) \cite{kessler2002short} along with personality information assessed using the Big Five model. K-10 is a distress scale and is used to evaluate depression and anxiety symptoms. Individuals were divided into two groups, "At-risk" and "No-risk," based on the K-10 scale, to assess depression risks. Those classified as "At-risk" have a K-10 score of 29 or more, while those classified as "No-risk" have a score of less than 20. Additional measures include musical engagement which describes music consumption behavior that is yet another indirect measure of well-being. Personality data and musical engagement are not used in the current analysis, but they were in the original study to assess internal consistency, which was found to be high.

\subsection{Multimodal Feature Extraction}
\label{section:feature interaction}

\subsubsection*{Sessions}
We used sessions as the basis for time resolution since we wanted to capture variations in users' temporal engagement. A session is defined as a period of continuous listening activity, surrounded by periods of inactivity of a minimum of two hours. The concept of sessions was used since people's preferences may differ during different sessions, and recommendations based on those may not be useful. \cite{inbook} Sessions with less than five songs were discarded. Further statistics such as the number of sessions, the number of unique songs, and the average length of a session are included in Table~\ref{table:data}.

\subsubsection*{Track Embeddings}
Each track\footnote{We use the terms song and track interchangeably.} in the session consists of user ID, song name, artist name, and time stamp. The listening history of all the users is used to prepare track embeddings, which are used as initialization in our architecture's embedding layer. Our approach for obtaining these is based on the CBOW model by Mikolov et al. \cite{mikolov2013efficient}. We try to employ a strategy similar to Word2Vec's by grouping songs that are frequently listened to together by users. The objective of this was to put songs that are likely to co-occur in a session together in an embedding space before passing them to the GRU network. A variety of models pursuing this approach have shown performance improvement \cite{cosernn}. We use the gensim \cite{rehurek_lrec} library with their default parameters for Word2Vec.

\subsubsection*{Acoustic Embeddings}
\begin{figure}[!t]
\centering
\includegraphics[width=\linewidth, keepaspectratio]{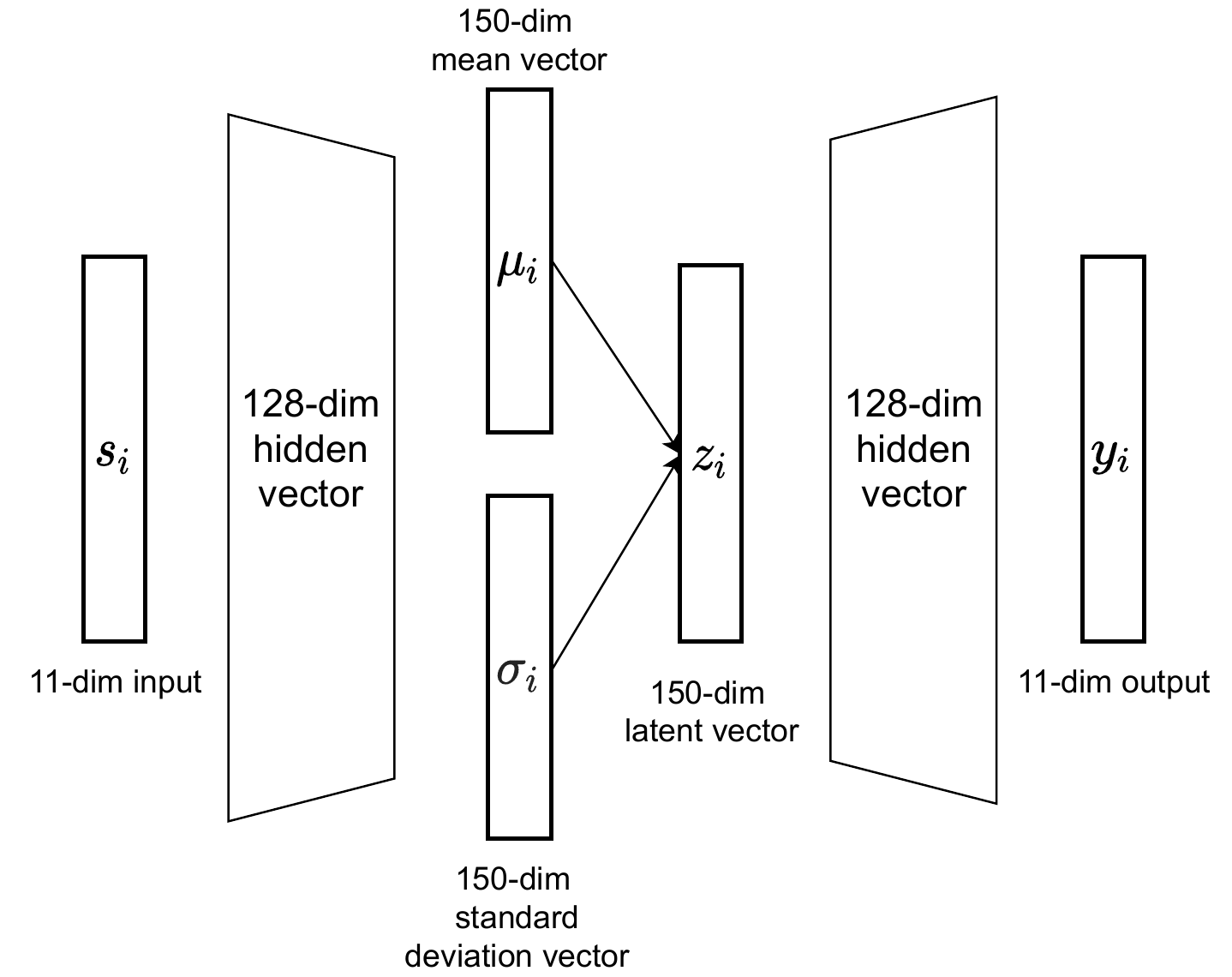}
\caption{VAE model for song features.}
\label{fig:vae}
\end{figure}

For each track, 11 audio features are obtained using the Spotify API. These features include Acousticness (probability that a track is acoustic); Danceability (how suitable a track is for dancing); Duration (of the track); Energy (perceptual measure of intensity and activity in a track); Instrumentalness (measure of a track containing no vocals); Liveness (probability that the track was performed in the presence of an audience); Loudness (average loudness of the track in decibel); Speechiness (presence of spoken words in a track); Tempo (pace of the track in beats per minute(BPM)); Valence (pleasantness conveyed by a track); Mode (major or minor). Instead of passing acoustic features to linear layers and learning the weights of the layers end-to-end, we generate latent vectors for them using an unsupervised learning task. A Variational Autoencoder (VAE) is used to extract acoustic embeddings. VAEs improve upon conventional autoencoders by producing a continuous latent space, and thus decoding any point from this space could create an acceptable representation that resembles the input \cite{kingma2013auto_vae}. The significance of this in our architecture is demonstrated in section~\ref{section:entire}. For any given track, we project its 11-dimensional feature vector $s_{i}$ to a 150-dimensional latent space $z_{i}$, which is used as acoustic embedding for the song. Let us establish some notation that could be used later. 

\begin{equation}
\label{equation:vae}
Encoder: z_{i} = G_{e}(s_{i}); \hspace{0.2cm} Decoder: ~y_{i} = G_{d}(z_{i})
\end{equation}
\\
The encoder and decoder used in Fig.~\ref{fig:vae} are represented using the function $G_{e}$ and $G_{d}$ respectively. A loss is computed for the reconstructed vector $y_{i}$ and the original vector $s_{i}$, which is used for training.

\subsubsection*{Lyrical Embeddings}
To obtain lyrical features, we first extract the lyrics of each track using the Genius API\footnote{https://www.genius.com}. The API needs to be constructed using the correct artist and track name, so we use a method of added web crawler to obtain the Genius website URL for the song's lyrics instead of hard-coding the artist and track name in Genius API. We use the Sentence-BERT \cite{sentencebert} model for the computation of lyrical embeddings. Sentence-BERT employs siamese and triplet networks to give semantically relevant sentence embeddings, in addition to the benefits of transformer models. 768-dimensional embeddings are produced for song lyrics, then streamlined to 150 using Principle Component Analysis. Due to certain songs having only acoustic content and others lacking lyrical data, lyrics were only accessible for roughly 80\% of the tracks. We pass zero-vector as embedding for songs without lyrics.

%dimensionality reduction technique
\subsubsection*{Tag Embeddings}
Last.fm tags were extracted via the Last.fm API for individual tracks. 300-dimensional FastText embeddings \cite{bojanowski2017enriching} were used, which were then reduced to 150 using linear layers for individual words in the tags. The embedding for all the words in the tags of the track was then averaged to create a song's tag embedding. Since tags were not available for all of the tracks in the dataset, we create a subset dataset with only those tracks with available tags and obtain the results using it. Table~\ref{table:data} shows a description of our dataset along with the mentioned subset dataset. 
%We did not propose our final model, depicted in Fig.~\ref{fig:model}, to be one with tags incorporated, for added modality, owing to the fact that the workable dataset shrinks a lot. Hence, we provide sufficient results which demonstrate the efficacy of the model with and without tags-based information in section~\ref{section:results}.
\begin{table}[h!]
\renewcommand{\arraystretch}{1.3}
\caption{Description of Dataset.}
\label{table:data}
\centering
% Some packages, such as MDW tools, offer better commands for making tables
% than the plain LaTeX2e tabular which is used here.
\begin{tabular}{|c|c|c|}
\hline
Explanation & Complete Dataset & Subset Dataset \\ \hline \hline
Number of Users       & 541              & 536            \\ \hline
Number of Sessions    & 394057             & 66803         \\ \hline
Unique Songs       & 413819             & 36661               \\ \hline
Logs (Total Listening Events)        & 5492139           & 463074               \\ \hline
Average Length of Session        &    13.93        &     6.93           \\ \hline
\end{tabular}
\end{table}

\begin{figure*}[h!]
\centering
% \fbox{\rule{0pt}{2in} \rule{.9\linewidth}{0pt}}
\includegraphics[width=0.9\linewidth,keepaspectratio]{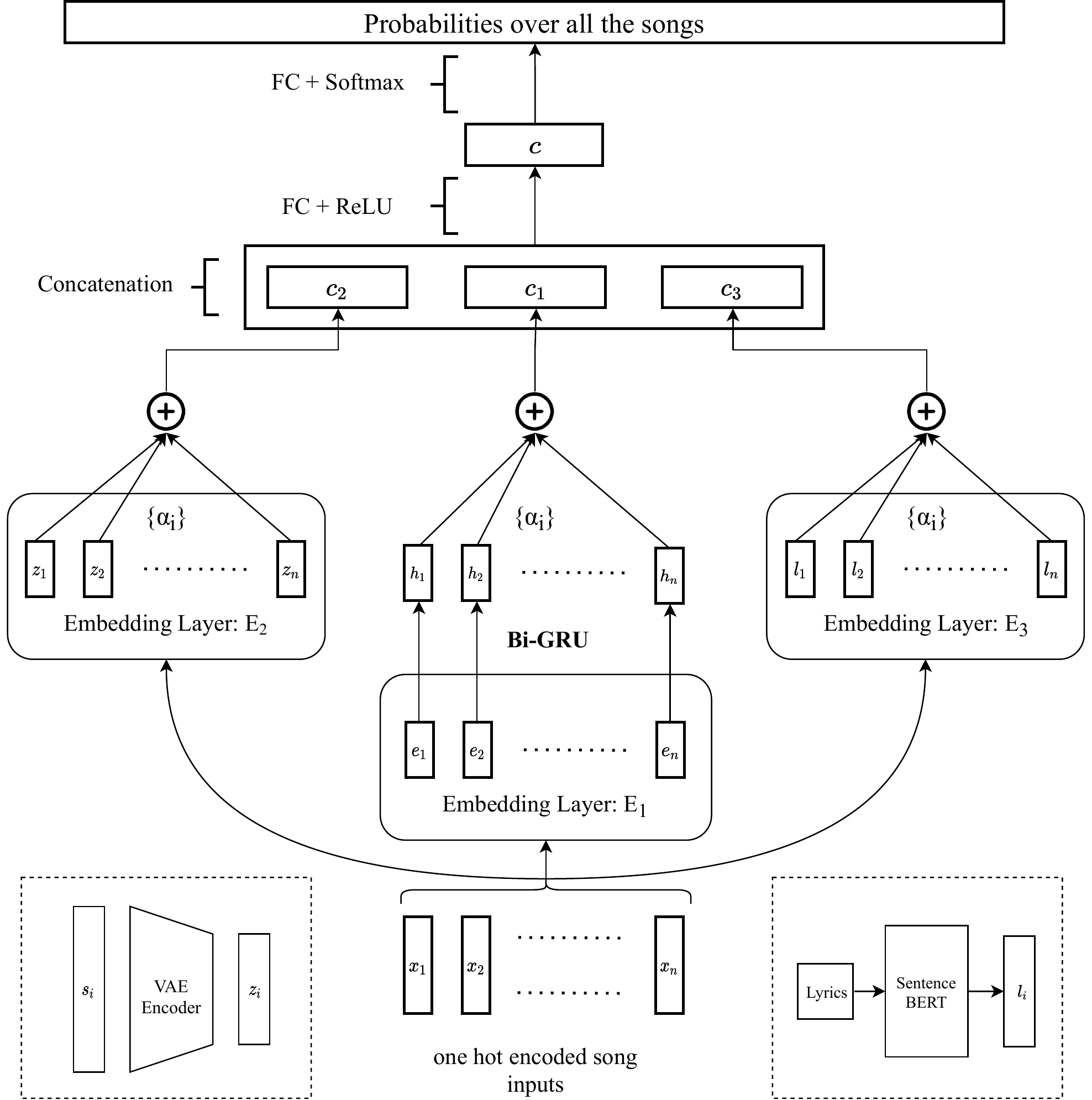}
\caption{Overview of our proposed music recommendations architecture (\textit{ANN-Word2Vec + Acoustic + Lyrics}). Given a series of tracks {$x_{i}$}, we first pass them through our first embedding layer {$E_{1}$} (track embeddings) which are then passed through BiGRU to produce hidden states {$h_{i}$}. Then, the attention layer generates attention weights {$\alpha_{i}$} using the hidden states. VAE Encoder and Sentence BERT models are used to generate representations from acoustic features ({$s_{i}$}) and Lyrics, which are used as initializations in Embedding layers {$E_{2}$} and {$E_{3}$} respectively. Concurrently, the sequence of tracks is passed from Embedding layers {$E_{2}$} (acoustic embeddings) and {$E_{3}$} (lyrical embeddings) to generate {$z_{i}$} and {$l_{i}$} respectively. Subsequently, individual context vectors are produced using the previously generated attention weights. The concatenated context vector is used for recommending the next track.}
\label{fig:model}
\end{figure*}

\subsection{Proposed Architecture}
\subsubsection*{Song Prediction Task}
Our models are fed a series of tracks in the order $\{x_{1}, x_{2}...x_{n}\}$, along with their attributes (lyrical, acoustic, tags), and our goal is to predict relevant songs for the user.

\label{section:entire}
Fig.~\ref{fig:model} illustrates our proposed architecture explained in this section (using lyrical and acoustic embeddings). The architecture starts with one hot encoding of tracks $\{x_{1}, x_{2}...x_{n}\}$, which are passed to our first embedding layer $E_{1}$. This embedding layer is initialized from the track embeddings $\{e_{1}, e_{2}...e_{n}\}$ obtained in section \ref{section:feature interaction}.

\begin{equation}
e_{i} = E_{1} * x_{i}
\end{equation}
\\
These are then passed to a BiGRU to produce two bidirectional hidden states (which we add to obtain a single vector $h$) for each time step. Additive attention \cite{bahdanau2016neural} is used on the hidden states to obtain a set of attention weights $\{\alpha_{1}, \alpha_{2}...\alpha_{n}\}$. We enable our first context vector to focus on sequential information by computing it from the weighted sum of hidden states using these weights.

\begin{equation}
h_{1}, h_{2}...h_{n} = BiGRU(e_{1}, e_{2}...e_{n}); \hspace{0.2cm} c_{1} = \sum_{i}\alpha_{i}h_{i}
\end{equation}
\\
Further, acoustic embeddings $\{z_{1}, z_{2}...z_{n}\}$ along with lyrical embeddings $\{l_{1}, l_{2}...l_{n}\}$ are incorporated into the architecture. These are used as initializations of our second ($E_{2}$) and third ($E_{3}$) embedding layers respectively.

\begin{equation}
z_{i} = E_{2} * x_{i}; \hspace{0.2cm} l_{i} = E_{3} * x_{i}
\end{equation}
\\
The initializations of all three embedding layers are fine-tuned during training. Here we argue the importance of using VAE, the new fine-tuned representations produced by the embedding layer would be perfectly valid due to being in a continuous space produced by the VAE \cite{kingma2013auto_vae}. Finally, separate context vectors are generated as a weighted sum of these features using the previously produced attention weights.

\begin{equation}
\label{equation:contextmodal}
c_{2} = \sum_{i}\alpha_{i}z_{i}; \hspace{0.2cm} c_{3} = \sum_{i}\alpha_{i}l_{i}
\end{equation}
\\
The context vectors are concatenated and then passed through a fully connected (FC) layer with a \textit{Leaky ReLU} activation to obtain a smaller dimensional vector $c$. Finally, an output vector with a size equal to the number of songs in the vocabulary is obtained by passing $c$ to another fully connected layer and a \textit{softmax} operation. This output vector gives us probabilities for all the tracks which are used for recommending the next track.

\begin{equation}
\label{equation:smallervector}
c = ReLU(FC([c_{1}; c_{2}; c_{3}])); \hspace{0.2cm} o = softmax(FC(c))
\end{equation}
\\
In the results, various combinations of modalities are examined. As a reference, we have created a set of notations for them here. The Attention architecture using only the context vector from the track embeddings (obtained using Word2Vec) is described as \textit{ANN-Word2Vec} (ANNW). Next, the architectures in which context vectors from acoustic and lyrical embeddings are combined individually are denoted as\textit{ ANNW + Acoustic} and \textit{ANNW + Lyrics}. Finally, we denote our complete architecture as \textit{ANNW + Acoustic + Lyrics}. We use the procedure described in \eqref{equation:contextmodal} to add another context vector for tag embeddings and concatenate it with the rest. This model's results are provided only for the subset dataset due to the unavailability of tags.

\subsection{Implementation Details}
\subsubsection*{Network training}
We use Nvidia’s GTX 1080Ti, with 11GB of VRAM to train our models. The embeddings, GRU hidden vectors, and latent vectors generated in VAEs each have a size of 150. The size of the smaller hidden vector produced in \eqref{equation:smallervector} was kept as 256. The size of one-hot encoded song inputs was kept equal to the number of tracks in our complete dataset. We use the Adam optimizer \cite{adam} with an initial learning rate of $1e^{-3}$ and Cross-Entropy Loss to train the complete architecture. A batch size of 32 and 0.2 dropout regularization was used for the embedding layers. PyTorch \cite{pytorch} was used for the implementation of the complete architecture.

\begin{table*}[!t]
\renewcommand{\arraystretch}{1.3}
\caption{Results: Comparison with state-of-the-art methods.}
\label{table:main_results}
\centering
% Some packages, such as MDW tools, offer better commands for making tables
% than the plain LaTeX2e tabular which is used here.
\begin{tabular}{|l|c|c|c|c|c|}
\hline
Methods & k=10 & k=20 & k=30 & k=40 & k=50 \\ \hline \hline
GRU4REC \cite{gru4rec}        & 15.78 &  18.20 & 19.58 & 20.53 & 21.27 \\ \hline
ANN \cite{sachdevaetal}            & 24.51 & 26.28 & 27.29 & 28.03 & 28.60 \\ \hline
ANN-LSA (Proposed Baseline)         & 27.37 & 29.34 & 30.46 & 31.26 & 31.87 \\ \hline
ANNW + Acoustic + Lyrics (Our Model) & \textbf{33.17} & \textbf{35.20} & \textbf{36.34} & \textbf{37.11} & \textbf{37.73} \\ \hline
\end{tabular}
\end{table*}

\subsubsection*{Evaluation Metrics}
To establish a fair comparison, we used the same measures as provided by Sachdeva et al. \cite{sachdevaetal}. The training data was formed from the first 70 percent of sessions for each user, in order of occurrence, and the remaining 30 percent was used for testing. We iterate through the listening histories of the users and use the next song for evaluation while giving songs till that point as input. The evaluation metric for all the models is HitRatio@k \cite{Lee2010}, where $k$ is the number of songs predicted and a hit is whether the required song is in the prediction set. We're attempting to predict top-$k$ items for users in the context of recommendation systems; thus we chose this metric to evaluate sessions obtained from individual user's listening histories.

The training sessions were used for pre-training track embeddings in section~\ref{section:feature interaction}. Random initialization is done for tracks not present in the training set. The VAE encoder is similarly trained using the acoustic features of all the tracks in the training set.

\section{Experiments \& Results}
In this section, we show a comprehensive evaluation of the proposed model and benchmark against recent state-of-the-art optimizations and deep learning-based algorithms. All of the trained models and code shall be made publicly available.

\subsection{Baselines}
The following is a list of the baseline models we choose to compare our model to, as well as the reasoning behind our decision.
\begin{enumerate}
   \item \textbf{GRU4REC:} We use the implementation of the model proposed by Hidasi et al. \cite{gru4rec}. Similar to our approach, they used GRU-based RNNs to model sessions of users' listening histories. The input to the model is the one-hot encoding of tracks.
   \item \textbf{ANN:} In line with the work done by  Sachdeva et al. \cite{sachdevaetal}, a base architecture was created using only sequential knowledge from user histories. The network's input is a one-hot encoding of the songs fed to an embedding layer, which generates song embeddings. Finally, these embeddings are transferred to a BiGRU and an attention layer on top of it to obtain a context vector. We adopt this as our baseline model since their work in using Attentive Networks for song predictions is close to ours.
   \item \textbf {ANN-LSA:} To contrast the impact of Word2Vec initializations, we compare it with another technique to create track embeddings. First, a session-track matrix is created for a combined set of all the sessions of each user. Each row indicates the frequencies of songs in that session. Owing to the sparsity of this matrix, we perform latent semantic analysis (LSA), which finds a low-rank approximation for the same using singular value decomposition. Finally, the column vectors of this reduced matrix are used as track embeddings and replaced with embeddings obtained by Word2Vec.
\end{enumerate}

\begin{table*}[h!]
\renewcommand{\arraystretch}{1.3}
\caption{\label{table:ablation_study}Ablation Study: Individual modalities in our model (\ref{section:entire}) are examined.}
\centering
% Some packages, such as MDW tools, offer better commands for making tables
% than the plain LaTeX2e tabular which is used here.
\begin{tabular}{|l|c|c|c|c|c|}
\hline
Methods & k=10 & k=20 & k=30 & k=40 & k=50 \\ \hline
\hline
\multicolumn{6}{|c|}{\textbf{Complete Dataset}}     \\ \hline
ANN             & 24.51 & 26.28 & 27.29 & 28.03 & 28.60 \\ \hline
ANN-LSA         & 27.37 & 29.34 & 30.46 & 31.26 & 31.87 \\ \hline
ANN-Word2Vec (ANNW)     & 29.68 & 31.98 & 33.33 & 34.26 & 35.01 \\ \hline
ANNW + Acoustic & 31.72 & 33.56 & 34.58 & 35.31 & 35.82 \\ \hline
ANNW + Lyrics  & 32.14 & 33.99 & 35.07 & 35.82 & 36.38 \\ \hline
ANNW + Acoustic + Lyrics & \textbf{33.17} & \textbf{35.20} & \textbf{36.34} & \textbf{37.11} & \textbf{37.73} \\ \hline
\hline
\multicolumn{6}{|c|}{\textbf{Subset Dataset}}       \\ \hline
ANN  & 32.81 & 33.97 & 34.72 & 35.31 & 35.84 \\ \hline
ANNW + Acoustic + Lyrics & 39.75 & 41.60 & 42.64 & 43.47 & 44.16 \\ \hline
ANNW + Acoustic + Lyrics + Tags & \textbf{40.72} & \textbf{42.50} & \textbf{43.50} & \textbf{44.27} & \textbf{44.93} \\ \hline
\end{tabular}
% \end{tabu}
% \end{adjustbox}
\end{table*}

\subsection{Results}
\label{section:results}

The results in Table~\ref{table:main_results} demonstrate the far superior performance of our method on the testing data when compared to studies that have attempted the same task. Our model shows 35\% improvement over the state-of-the-art ANN approach \cite{sachdevaetal}. The numbers in the table correspond to our evaluation metric (HitRatio@k), which shows that as k increases, the model's prediction accuracy improves due to larger prediction sets.

% \begin{table}[h!]
% \begin{center}
% \caption{User Study.}
% \setlength{\tabcolsep}{8pt}
% \begin{tabular}{|l|l|l|l|l|l|}
% \hline
% Methods & k=10  & k=20  & k=30  & k=40  & k=50  \\ \hline \hline
% At-risk-142 & 32.49 & 34.23 & 35.12 & 35.78 & 36.23 \\ \hline
% No-risk-193 & 28.54 & 29.83 & 30.50 & 30.99 & 31.40 \\ \hline
% No-risk-142 & 28.43 & 29.73 & 30.44 & 30.92 & 31.29 \\ \hline
% \end{tabular}
% \end{center}
% % \end{tabu}
% % \end{adjustbox}
% \end{table}

\subsection{Ablation Study}
We investigate several combinations of the modalities to gather a better understanding of our proposed architecture. The results of the experiments on our evaluation metric (HitRatio@k) can be found in Table~\ref{table:ablation_study}.

\subsubsection*{Complete Dataset}
To begin with, \textit{ANNW}'s better performance in producing track embeddings justifies its advantage over other techniques such as LSA. This can be reasoned out by the fact that Word2Vec embeddings used in \textit{ANNW} naturally preserve more information about tracks near to them, which is more suitable for a sequential music recommendation task. Following that, we present individual findings for lyrical and acoustic modalities, with lyrics outperforming acoustic modalities. The rise in the accuracy of individual models demonstrates the importance of integrating modalities in sequential music recommendation models since modalities encapsulate additional information absent from song embedding structures. Finally, the performance of models using individual modalities was inferior to a fusion model that incorporates all of them. This illustrates that, while lyrical and acoustic modalities might have some overlapping information, they also provide complementary information which should be used in fusion models.

\subsubsection*{Subset Dataset}
The experiments were also conducted on the subset dataset to fuse tag-based modality. Incorporating yet another modality yielded even better results. We did not propose our final model, depicted in Fig.~\ref{fig:model}, to be one with tags incorporated, for added modality, owing to the fact that the workable dataset shrinks a lot. Hence, we provide sufficient results which demonstrate the efficacy of the model with and without tags-based information.

\subsection{User Study}

We perform a study to see the influence of users' psychological health on the performance of our model; for this, we divide users into two groups: At-risk and No-risk. Following the approach used in Surana et al. \cite{suranaetal}, users with K-10 score less than 20 were classified as No-risk, while those with K-10 score greater than 29 were classified as At-risk. This resulted in 142 At-risk users and 193 No-risk users. To provide a balanced comparison, we further show evaluations on the 142 No-risk users sorted in decreasing order of K-10 values. Owing to its superior performance, the model (\textit{ANNW + Acoustic + Lyrics}) was used to train each set of users. 

\begin{table}[h!]
\renewcommand{\arraystretch}{1.3}
\caption{User Study: Groups based on Kessler's Distress Scale.}
\label{table:user}
\centering
% Some packages, such as MDW tools, offer better commands for making tables
% than the plain LaTeX2e tabular which is used here.
\begin{tabular}{|l|l|l|l|l|l|}
\hline
Methods & k=10  & k=20  & k=30  & k=40  & k=50  \\ \hline \hline
At-risk-142 & 32.49 & 34.23 & 35.12 & 35.78 & 36.23 \\ \hline
No-risk-193 & 28.54 & 29.83 & 30.50 & 30.99 & 31.40 \\ \hline
No-risk-142 & 28.43 & 29.73 & 30.44 & 30.92 & 31.29 \\ \hline
\end{tabular}
% \end{tabu}
% \end{adjustbox}
\end{table}

The results for the performed study can be found in Table~\ref{table:user}. As hypothesized, the model trained on At-risk users demonstrates higher prediction accuracy than both sets of No-risk users. This highlights the disparity in listening behavior between the two groups; future research might look for more definitive trends in this direction.

% \begin{table}[h!]
% \begin{center}
% \caption{\label{table:ablation_study}Ablation Study: Individual modalities in our model (\ref{section:entire}) are examined.}
% \setlength{\tabcolsep}{4pt}
% \begin{tabular}{|l|c|c|c|c|c|}
% \hline
% Methods & k=10 & k=20 & k=30 & k=40 & k=50 \\ \hline
% \hline
% \multicolumn{6}{|c|}{\textbf{Complete Dataset}}     \\ \hline
% ANN             & 24.51 & 26.28 & 27.29 & 28.03 & 28.60 \\ \hline
% ANN-LSA         & 27.37 & 29.34 & 30.46 & 31.26 & 31.87 \\ \hline
% ANN-Word2Vec (ANNW)     & 29.68 & 31.98 & 33.33 & 34.26 & 35.01 \\ \hline
% ANNW + Acoustic & 31.72 & 33.56 & 34.58 & 35.31 & 35.82 \\ \hline
% ANNW + Lyrics  & 32.14 & 33.99 & 35.07 & 35.82 & 36.38 \\ \hline
% ANNW + Acoustic + Lyrics & \textbf{33.17} & \textbf{35.20} & \textbf{36.34} & \textbf{37.11} & \textbf{37.73} \\ \hline
% \hline
% \multicolumn{6}{|c|}{\textbf{Subset Dataset}}       \\ \hline
% ANN  & 32.81 & 33.97 & 34.72 & 35.31 & 35.84 \\ \hline
% ANNW + Acoustic + Lyrics & 39.75 & 41.60 & 42.64 & 43.47 & 44.16 \\ \hline
% ANNW + Acoustic + Lyrics + Tags & \textbf{40.72} & \textbf{42.50} & \textbf{43.50} & \textbf{44.27} & \textbf{44.93} \\ \hline
% \end{tabular}
% \end{center}
% \end{table}

\section{Conclusion}
Our work in this paper focuses on using multiple modalities for sequential music recommendations and outperforms existing state-of-the-art models. While using embeddings from a single modality, results demonstrate that using lyrics alone performs better than other modalities. Fusing modalities proves to be the best, as demonstrated by the performance of the \textit{ANNW + Acoustic + Lyrics + Tags} model, albeit on the subset dataset. There are certain limitations to designing a multimodal RS. Firstly, although majority of the tracks in an individual's listening history comprise lyrics, not all contain lyrics. Despite this, our approach demonstrates superior performance. Another limitation is that user-defined tags are not available for every track. One way to circumvent this is to use NLP techniques such as Topic Modelling, Sentiment Analysis, to extract relevant information from lyrics. This further underlines the importance of incorporating lyrics in designing Music RS. Furthermore, using acoustic features to identify genres and emotions can provide additional tags in cases of missing lyrics and tags. In the future, it remains to be seen if tags generated via lyrics or acoustic features give comparable results. Finally, our case study on users categorized as At-risk for depression highlights the importance of individual differences in online music consumption. The higher accuracy can be attributed to repetitive listening behavior associated with the At-risk group, as demonstrated by several studies \cite{suranaetal,Saarikallio}. This emphasizes the importance of integrating user states and traits into recommendation systems in order to create more personalized recommendations.

\bibliographystyle{IEEEtran}
\bibliography{IEEEabrv,mybibliography}

\end{document}